\documentclass[a4paper,12pt]{article}
\usepackage[T1]{fontenc}
\usepackage{graphicx}
\usepackage{color}

\usepackage{amsmath,amssymb,amsthm,ascmac,authblk,bm,booktabs,caption,comment,enumerate,latexsym,lscape,longtable,mathabx,mathrsfs,mathtools,multirow,natbib,newtxtext,psfrag,setspace,subcaption,thmtools,paralist,enumitem}
\usepackage{array,tabularx,threeparttable}
\usepackage{placeins}
\usepackage[figuresright]{rotating}
\usepackage[stable]{footmisc}
\definecolor{mycolor}{cmyk}{0.85,0.21,0,0.06}
\definecolor{applegreen}{rgb}{0.55,0.71,0.0}
\definecolor{blush}{rgb}{0.87,0.36,0.51}
\usepackage[colorlinks=true,linkcolor=mycolor,filecolor=mycolor,urlcolor=mycolor,citecolor=mycolor,backref=false,driverfallback=dvipdfmx]{hyperref}
\usepackage[margin=2.3cm]{geometry}
\usepackage{chngcntr}
\usepackage{xr}
\usepackage{clipboard}
\usepackage{microtype}

\newcolumntype{Y}{>{\raggedright\arraybackslash}X}
\setcitestyle{authoryear,round,aysep={,},yysep={,},citesep={;}}

\renewcommand{\hat}{\widehat}

\allowdisplaybreaks[1]

\theoremstyle{definition}

\newtheorem{proposition}{Proposition}

\numberwithin{equation}{section}

\definecolor{myred}{rgb}{0.76,0.23,0.13}
\definecolor{myblue}{rgb}{0.286,0.427,0.714}
\definecolor{mygreen}{rgb}{0.00,0.45,0.00}

\begin{document}

\title{\bfseries Local and Neighboring Food Retail Supply and Population across Japan's Municipalities}
\author{Lei Qin\vspace{-0.35cm}\thanks{471 Ebigase, Higashi-ku, Niigata, 950-8680, Japan. Email: \href{mailto:qinlei6d@unii.ac.jp}{qinlei6d@unii.ac.jp}}\\
    Faculty of International Economic Studies, University of Niigata Prefecture\\
    }
\date{\today}
\pagestyle{plain}
\maketitle

\begin{abstract}
This paper examines the relationship between food retail supply and resident population across Japanese municipalities. 
I develop a partial-equilibrium framework in which local food retail improves consumption convenience and population density supports retail entry. 
The framework allows positive local feedback and distinguishes local retail from neighboring consumption opportunities. 
Using Japan's 2010, 2015, and 2020 Population Censuses and the 2012, 2016, and 2021 Economic Censuses, I estimate municipality fixed-effects specifications with spatially lagged explanatory variables. 
Local and neighboring food and beverage retail supply (FRS) are positively associated with resident population.
The corresponding coefficients are 0.0893 and 0.1364, respectively, and remain positive when prefecture-by-wave fixed effects are introduced. 
A common-sample comparison shows that the standardized local FRS association is stronger than that of food-service supply (FSS). 
The estimated local FRS relationship also varies across retail scale, although the magnitude of this nonlinearity and the negative slope at high establishment counts are sensitive to regional time effects and the estimation period. 

\bigskip


\bigskip

\end{abstract}

\vfill

\newpage
\doublespacing

\section{Introduction}\label{sec:introduction}

Food is a necessity, but the convenience of obtaining food varies across locations. 
Even when food products can be transported across regions, consumers depend on nearby stores, shopping time, and local services to access them. 
This creates a two-way relationship between retail supply and residential location. 
A larger population supports retail establishments through local demand, while better retail access can make a municipality more attractive to residents. 
Yet these relationships do not necessarily follow administrative boundaries. 
A municipality with few stores may still benefit from extensive retail opportunities in neighboring areas, which can also compete with its own establishments. 
Thus, understanding the relationship between food retail and population requires examining both local and neighboring markets. 

I develop a partial-equilibrium framework that links food access, residential location, and retail entry. 
Food retail raises residents' consumption surplus by reducing access costs and providing convenience and variety. 
Housing adjustment translates this surplus into population density, while retail establishments enter in response to local demand and operating costs. 
The model yields three implications. 
First, population and retail supply can reinforce each other when the demand effect of population density exceeds its effect on operating costs and the equilibrium is locally stable. 
Second, diminishing marginal benefits from food access weaken the population response to improvements in retail supply. 
Third, neighboring retail provides additional consumption opportunities but may displace local retail demand, making its total equilibrium effect ambiguous. 
I examine these relationships using Japanese municipality-level Census data and fixed-effects regressions with spatially lagged explanatory variables. 

The empirical contribution has three parts. 
(\romannumeral1) It documents the conditional relationship between resident population and local and neighboring food and beverage retail supply (FRS) in a national municipality panel for Japan. 
This distinction allows retail markets to extend beyond administrative boundaries without treating neighboring establishments as equivalent to local establishments. 
(\romannumeral2) It compares FRS and food-service supply (FSS) using the same municipalities, fixed effects, and spatial specification. 
The comparison identifies a stronger standardized local population association for FRS. 
(\romannumeral3) It examines whether the local FRS relationship varies across retail scale. 

I combine the 2012, 2016, and 2021 Economic Censuses with the 2010, 2015, and 2020 Population Censuses on harmonized municipal boundaries. 
The resulting balanced panel contains 1,725 municipalities and 5,175 observations. 
Population observations precede their paired establishment observations by 8 to 16 months. 
The empirical specifications relate log resident population to transformed local establishment counts and spatially weighted neighboring counts. 
Municipality fixed effects absorb time-invariant municipal characteristics, while wave and prefecture-by-wave fixed effects account for common and region-specific time variation. 

The linear specification yields positive coefficients on both local and neighboring FRS, 0.0893 and 0.1364, respectively. 
These associations remain similar after introducing prefecture-by-wave fixed effects. 
In the joint FRS--FSS specification, the local FRS coefficient is 0.0928, while the local FSS coefficient is $-0.0052$ and imprecisely estimated. 
The standardized difference between the two local coefficients is positive, with a 95\% confidence interval that excludes zero. 

The nonlinear estimates also vary across retail scale and model specifications. 
The benchmark quadratic model yields a declining fitted local slope, but the estimated slope at the 90th percentile changes from $-0.0841$ to approximately zero when prefecture-by-wave fixed effects are introduced. 
The estimated curvature also weakens substantially when the final observation period is excluded. 
Thus, the evidence supports scale-dependent heterogeneity in the conditional relationship, but not a robust negative population association at high retail levels. 

The paper relates first to the urban and regional literature on local amenities and residential location. 
In the spatial-equilibrium tradition, local amenities enter households' location choices together with wages and housing costs \citep{roback1982wages,glaeser2009wealth}. 
Amenities can also help explain within-city location patterns \citep{brueckner1999central}. 
The consumer-city literature emphasizes that dense places provide consumption opportunities that can contribute to urban attractiveness and growth \citep{glaeser2001consumer}. 
More recent work shows that endogenous amenities can reinforce residential sorting \citep{diamond2016determinants} and that nontradable services can influence residential location within cities \citep{couture2020urban}. 
\citet{DoddsDubrovinsky2015} model the interaction between retail amenities and population density, showing how retail agglomeration can shape urban spatial equilibria. 
\citet{AlmagroDominguezIino2025} jointly model residential sorting and the endogenous provision of nontradable consumption amenities. 
Using evidence from Amsterdam, they show how differences in local demand composition influence the supply of different amenities. 
These studies establish a theoretical and empirical basis for treating local services and residential location as jointly determined outcomes.

A second literature studies how local market size shapes the supply and variety of private goods. 
When provision involves fixed costs and markets are geographically limited, the composition of local demand affects which private goods can be supported \citep{waldfogel2008median}. 
Using restaurants, \citet{schiff2015cities} shows that larger and more spatially concentrated markets sustain greater product variety. 
For grocery products, \citet{handbury2015goods} show that product availability itself varies systematically across cities and matters for measured local price levels. 
These studies demonstrate how local demand can shape the availability and variety of private consumption services. 
They also highlight the possibility of reverse causality in the relationship between population and retail establishments. 

A third literature examines food availability and spatial access more directly. 
\citet{Nilsson2023} examines the relationship between population density and food-store proximity using Swedish neighborhood panel data. 
The study finds that increases in population density are associated with shorter distances to the nearest food store, particularly in rural areas, with the relationship driven mainly by growing regions. 
It focuses on the relationship between local market size and store proximity, while acknowledging that retail access and residential location may be jointly determined. 
Using Swedish village-shop data, \citet{AmcoffMollerWestholm2011} find no significant change in local in- or out-migration associated with shop closures. 
\citet{allcott2019food} distinguish retail environments from household demand when studying food deserts and nutritional inequality, illustrating why local supply and household outcomes are separate empirical objects. 
In Japan, the Ministry of Agriculture, Forestry and Fisheries' food-access work explicitly combines store proximity with residents' mobility constraints rather than treating establishment counts alone as complete access \citep{primaff2024access}. 
Official household-expenditure tabulations also show that purchases regularly cross municipal boundaries, particularly in smaller municipalities \citep{statjapan2011purchase}. 
This motivates separating establishments inside a municipality from those in neighboring municipalities. 

Finally, the empirical specification relates to spatial econometric models that distinguish local explanatory variables from their spatially lagged counterparts. 
The spatial lag of explanatory variables (SLX) specification includes neighboring retail conditions without imposing an endogenous spatial lag of population \citep{lesage2009introduction,halleckvega2015slx}. 

\section{Theoretical Framework}\label{sec:framework}

\subsection{Food Access Surplus}\label{subsec:food_access_surplus}

Let $x_m = \log(1 + R_m)$ denote local FRS, where $R_m$ is the number of food-retail establishments. 
Let $e_m$ denote neighboring FRS. 
Residents consume a standardized food quantity $q$ subject to an upper bound $\bar q$. 
The upper bound captures the feature of food as a necessity that food demand cannot expand without limit. 
Let $u(q)$ denote utility from food quantity, with $u'(q) > 0$ and $u''(q) < 0$ over its support. 
Let $a_m(x_m, e_m)$ and $v_m(x_m, e_m)$ denote the non-pecuniary cost of obtaining food and the convenience and variety services, respectively. 
The indirect surplus from food access is
\begin{equation}
\label{eq:food_access_surplus}
B_m(x_m, e_m) = \max_{0 \leq q \leq \bar q} \left\{ u(q) - [p_m + a_m(x_m, e_m)] q + \chi_m v_m(x_m, e_m) \right\}, 
\end{equation}
where $p_m$ is the monetary food price and $\chi_m \geq 0$ is the value placed on convenience. 
Assume the optimal food quantity is positive, $q_m^* > 0$. 
Assume $\partial a_m / \partial x_m < 0$, $\partial a_m / \partial e_m < 0$, $\partial v_m / \partial x_m > 0$, and $\partial v_m / \partial e_m \geq 0$ over the relevant support. 
Under a unique optimum and a fixed feasible set, the envelope theorem gives 
\begin{equation}
\label{eq:food_access_derivative_local}
\frac{\partial B_m(x_m, e_m)}{\partial x_m} = -\frac{\partial a_m(x_m, e_m)}{\partial x_m} q_m^* + \chi_m\frac{\partial v_m(x_m, e_m)}{\partial x_m} > 0, 
\end{equation}
and 
\begin{equation}
\label{eq:food_access_derivative_neighbor}
\frac{\partial B_m(x_m, e_m)}{\partial e_m} = -\frac{\partial a_m(x_m, e_m)}{\partial e_m} q_m^* + \chi_m \frac{\partial v_m(x_m, e_m)}{\partial e_m} \geq 0. 
\end{equation}
Thus, local and nearby retail can raise location utility by reducing access costs and expanding the local consumption set. 
This result is consistent with the broader consumption-amenity literature \citep{glaeser2001consumer,couture2020urban}. 

The quantity ceiling matters for curvature. 
When the quantity ceiling is strictly binding, $q_m^* = \bar q$ and $\partial q_m^* / \partial x_m = 0$. 
If $\partial^2 a_m / \partial x_m^2 > 0$ and $\partial^2 v_m / \partial x_m^2 < 0$, then 
\begin{equation}
\label{eq:food_access_concavity_satiated}
\frac{\partial^2 B_m(x_m, e_m)}{\partial x_m^2} = - \frac{\partial^2 a_m(x_m, e_m)}{\partial x_m^2} \bar q + \chi_m \frac{\partial^2 v_m(x_m, e_m)}{\partial x_m^2} < 0. 
\end{equation}
When the quantity ceiling is slack, however, a reduction in access costs can also increase food quantity. 
Thus, concavity is not an automatic implication. 

\subsection{Location Choice, Housing Adjustment, and Population Density}\label{subsec:population_density_model}

After solving the food subproblem, a resident consumes numeraire $c_m^0$ and housing $h_m$.
Utility and the housing budget are
\begin{equation}
\label{eq:location_utility}
U_m = c_m^0 + \alpha \log h_m + B_m(x_m, e_m) + A_m^O, \qquad \alpha > 0 
\end{equation}
and 
\begin{equation}
\label{eq:housing_budget}
c_m^0 + r_m h_m = y_m, 
\end{equation}
respectively, where $A_m^O$ collects other local amenities, $r_m$ is housing rent, and $y_m$ is income. 
Housing demand is $h_m^* = \alpha / r_m$. 
With outside utility $\bar V$, the location condition $V_m = \bar V$ implies 
\begin{equation}
\label{eq:bid_rent}
\log r_m = \frac{1}{\alpha} \left[ y_m + A_m^O + B_m(x_m, e_m) + \alpha \log \alpha - \alpha - \bar V \right]. 
\end{equation}
This is the standard local-amenity representation where a location that provides greater consumption surplus can support a higher bid rent \citep{roback1982wages,brueckner1999central,diamond2016determinants}. 

Let developers choose floor-area ratio $\rho_m$ with cost
\begin{equation}
\label{eq:housing_cost}
K(\rho_m)=\lambda_H\rho_m^\gamma, 
\qquad
\lambda_H>0, 
\quad
\gamma>1. 
\end{equation}
The first-order condition is
\begin{equation}
\label{eq:housing_supply_foc}
r_m=\lambda_H\gamma\rho_m^{\gamma-1}.
\end{equation}
Population density equals supplied housing area divided by housing demand, $n_m=\rho_m/h_m^*$.
Combining Equations~\eqref{eq:bid_rent}--\eqref{eq:housing_supply_foc} gives
\begin{equation}
\label{eq:population_density_equation}
d_m
\equiv
\log n_m
=
\bar d_m+\omega B_m(x_m,e_m),
\end{equation}
where
\begin{equation}
\label{eq:omega}
\omega
=
\frac{\gamma}{\alpha(\gamma-1)}>0.
\end{equation}
The term $\bar d_m$ collects income, outside utility, and constants that do not vary with food retail in this partial-equilibrium comparison. 
Hence, we have 
\begin{equation}
\label{eq:demand_side_slope}
\frac{\partial d_m}{\partial x_m} = \omega \frac{\partial B_m(x_m,e_m)}{\partial x_m}>0, 
\end{equation}
and 
\begin{equation}
\label{eq:demand_side_slope_neighbor}
\frac{\partial d_m}{\partial e_m} = \omega \frac{\partial B_m(x_m, e_m)}{\partial e_m} \geq 0. 
\end{equation}

\begin{proposition}[Food access and population density]
Holding income, other amenities, outside utility, housing-production conditions, and neighboring FRS fixed, greater local FRS raises equilibrium population density whenever $\partial B_m/\partial x_m>0$. 
Holding local FRS fixed, greater neighboring FRS raises population density through the direct food-access channel whenever $\partial B_m/\partial e_m>0$.
\end{proposition}

\subsection{Retail Entry and Local Feedback}\label{subsec:retail_entry_feedback}

The reverse channel operates through local demand.
A larger population density expands the customer base, while high-density locations can also have higher wages, commercial rents, and other operating costs.
Let a representative establishment's operating profit be
\begin{equation}
\label{eq:retail_profit}
\pi_m = \Omega \exp \left[ (\eta_D - \eta_C) d_m + z_m - \nu x_m \right] - K_R, 
\end{equation}
where $\Omega > 0$ and $K_R > 0$ are baseline gross-operating-surplus scale and fixed operating cost of an establishment, respectively. 
The parameters $\eta_D>0$ and $\eta_C\geq0$ measure the demand and local-cost effects of population density, respectively. 
The variables $z_m$ and $\nu>0$ represent retail productivity and competition among establishments, respectively. 
The exponential ensures that gross operating surplus is positive and makes its logarithm linear in density, the supply shifter, and retail scale. 
With interior free entry, $\pi_m = 0$, so 
\begin{equation}
\label{eq:retail_supply_equation}
x_m = \bar x_m + \xi d_m + \psi z_m, 
\end{equation}
where
\begin{equation}
\label{eq:retail_supply_parameters}
\xi = \frac{\eta_D - \eta_C}{\nu},
\qquad
\psi = \frac{1}{\nu} > 0. 
\end{equation}
The intercept is $\bar x_m = \log(\Omega / K_R) / \nu$ under this baseline specification. 
Thus, a positive density-to-retail channel requires $\eta_D > \eta_C$. 
This fixed-cost mechanism is closely related to evidence that local population and spatial concentration support differentiated local services \citep{waldfogel2008median,schiff2015cities}. 

For given neighboring supply, equilibrium is obtained by combining the population-density relation in Equation~\eqref{eq:population_density_equation} with the retail-entry relation in Equation~\eqref{eq:retail_supply_equation}.
Define the local feedback coefficient
\begin{equation}
\label{eq:feedback_coefficient}
\Lambda_m
=
\xi\omega
\frac{\partial B_m(x_m^*, e_m)}{\partial x_m}.
\end{equation}
If $\partial B_m / \partial x_m > 0$, $\xi > 0$, and
\begin{equation}
\label{eq:local_stability}
0 < \Lambda_m < 1,
\end{equation}
then a supply-side improvement $z_m$ raises both equilibrium retail supply and population density. 
I define Equation \eqref{eq:local_stability} as the \emph{local-stability condition} because the positive feedback between population density and retail supply converges locally under successive adjustments of the two theoretical relations. 
The comparative statics are
\begin{equation}
\label{eq:equilibrium_retail_response}
\frac{\partial x_m^*}{\partial z_m} = \frac{\psi}{1 - \Lambda_m} > 0, 
\end{equation}
\begin{equation}
\label{eq:equilibrium_population_response}
\frac{\partial d_m^*}{\partial z_m} = \frac{\omega \left[ \partial B_m(x_m^*, e_m) / \partial x_m \right] \psi}{1 - \Lambda_m} > 0. 
\end{equation}

\begin{proposition}[Local feedback and amplification]
Suppose $\eta_D > \eta_C$ and $0 < \Lambda_m < 1$. 
A favorable local retail-supply shifter raises both FRS and population density. 
Both equilibrium responses contain the amplification factor $1 / (1 - \Lambda_m) > 1$. 
\end{proposition}

\subsection{Nonlinearity and Neighboring Supply}\label{subsec:theory_nonlinearity_neighbor}

The model distinguishes the direct relationship between local retail and population density from the equilibrium response to a retail-supply shifter.
Holding neighboring supply and the other location attributes fixed, Equation \eqref{eq:demand_side_slope} gives the direct slope $\partial d_m / \partial x_m = \omega \, \partial B_m / \partial x_m$. 
If the marginal food-access surplus $\partial B_m / \partial x_m$ decreases with local retail scale, this slope decreases as well. 

To express that response as a function of the equilibrium level of local FRS, define
\begin{equation}
\label{eq:population_multiplier}
M_z(x_m) = \frac{\psi \omega \left[ \partial B_m(x_m, e_m) / \partial x_m \right]}{1 - \xi \omega \left[ \partial B_m(x_m, e_m) / \partial x_m \right]}
\end{equation}
as the response of log population density to the supply shifter. 
Thus, $M_z(x_m^*) = \partial d_m^*/\partial z_m$ at a locally stable equilibrium. 
Differentiation with respect to $x_m$ yields
\begin{equation*}
\frac{\partial M_z(x_m)}{\partial x_m} = \frac{\psi \omega \left[ \partial^2 B_m(x_m, e_m) / \partial x_m^2 \right]}{\left\{ 1 - \xi \omega \left[ \partial B_m(x_m, e_m) / \partial x_m \right] \right\}^2} \leq 0 
\end{equation*}
whenever food-access surplus is concave in $x_m$ and the denominator is nonzero. 
Under these conditions, diminishing marginal food-access surplus reduces both the direct population-density slope and the equilibrium response to a favorable retail-supply shifter. 
Both responses remain nonnegative when the direct access benefit is nonnegative and the equilibrium is locally stable. 

The net population response to neighboring FRS also depends on whether residents' use of neighboring stores displaces local retail demand. 
Equation \eqref{eq:food_access_derivative_neighbor} captures the direct access benefit of neighboring FRS when local FRS is held fixed. 
To allow displacement, extend the baseline entry relation in Equation \eqref{eq:retail_supply_equation} by adding a neighboring-retail term: 
\begin{equation}
\label{eq:neighboring_retail_supply}
x_m = \bar x_m + \xi d_m + \psi z_m - \zeta e_m, 
\qquad 
\zeta \geq 0. 
\end{equation}
This is an additional modeling assumption compared to Equation \eqref{eq:retail_supply_equation}, where $\zeta = 0$ recovers the baseline. 
Hence, a rise in $e_m$ can increase food-access surplus directly while reducing the number of local establishments supported by demand. 
Differentiating the population relation gives $\partial d_m^* / \partial e_m = \omega[(\partial B_m / \partial x_m)(\partial x_m^* / \partial e_m) + \partial B_m / \partial e_m]$. 
The augmented entry relation gives $\partial x_m^* / \partial e_m = \xi (\partial d_m^* / \partial e_m) - \zeta$. 
Combining them gives 
\begin{equation}
\label{eq:neighboring_population_response}
\frac{\partial d_m^*}{\partial e_m} = \frac{\omega \left[\partial B_m / \partial e_m - \zeta \left(\partial B_m / \partial x_m \right) \right]}
{1 - \xi \omega \left(\partial B_m / \partial x_m \right)}. 
\end{equation}
The derivatives of $B_m$ are evaluated at $(x_m^*, e_m)$. 
Under local stability, the denominator is positive, but the numerator combines a nonnegative direct access term with a non-positive substitution term. 

\begin{proposition}[Nonlinearity and neighboring supply] \label{pro:nonlinearity}
If the marginal food-access surplus from local FRS declines with retail scale, the model predicts a declining direct population--FRS slope and a weaker local feedback multiplier at higher FRS. 
By contrast, the total population response to neighboring FRS is ambiguous because the direct access benefit and the substitution away from local establishments work in opposite directions. 
\end{proposition}

\section{Data and Empirical Design}\label{sec:empirical_relations}

\subsection{Data and Spatial Coverage}\label{subsec:data_spatial_coverage}

The unit of observation is a municipality observed in three paired Census waves. 
Resident population is drawn from Japan's 2010, 2015, and 2020 Population Censuses, published by the Statistics Bureau of Japan. 
Establishment counts come from the 2012, 2016, and 2021 Economic Censuses for Business Activity, conducted by the Ministry of Internal Affairs and Communications and the Ministry of Economy, Trade and Industry. 
I pair population in 2010, 2015, and 2020 with establishments in 2012, 2016, and 2021, respectively. 
All three waves are aligned to fixed 2020 municipal boundaries so that the municipality identifier refers to the same geographic unit over time. 

FRS records the number of establishments in broad food and beverage retail, while FSS records establishments providing meal-oriented food services. 
Table \ref{tab:industry_scope} reports the official industry categories used to construct these establishment counts. 
FRS includes liquor retailers, and excludes general merchandise retailers classified under industry group 56, including department stores and general supermarkets. 
The 2012 Census uses the 2007 industry classification, while the 2016 and 2021 Censuses use its 2013 revision. 

\begin{table}[t]
\centering
\begin{threeparttable}
\caption{Broad official industry scope for FRS and meal-oriented FSS}
\label{tab:industry_scope}
\normalsize
\setlength{\tabcolsep}{5pt}
\renewcommand{\arraystretch}{1.08}
\begin{tabularx}{\textwidth}{@{}lY@{\hspace{1.5em}}lY@{}}
\toprule
\multicolumn{2}{c}{Food and beverage retail supply (FRS)} &
\multicolumn{2}{c}{Meal-oriented food-service supply (FSS)} \\
\cmidrule(lr){1-2}\cmidrule(lr){3-4}
Code & Industry category & Code & Industry category \\
\midrule
581 & Grocery stores & 761 & Eating and drinking places, except specialty restaurants \\
582 & Fruit and vegetable retailers & 762 & Specialty restaurants \\
583 & Meat and poultry stores & 763 & Soba and udon restaurants \\
584 & Fresh fish retailers & 764 & Sushi restaurants \\
585 & Liquor retailers & 767 & Coffee shops \\
586 & Confectionery and bakery retailers & 769 & Other eating and drinking places \\
589 & Retail food and beverage stores, n.e.c. & 771 & Takeout food services \\
    &                                    & 772 & Food delivery services \\
\bottomrule
\end{tabularx}
\begin{tablenotes}[flushleft]
\small
\item \textit{Notes:} Drinking houses and beer halls (765), bars, cabarets and nightclubs (766), and administrative or auxiliary activity rows (580, 760, and 770) are excluded. 
When code 589 is used, its subordinate recap rows 58A and 58B are not added separately. Codes 771 and 772 are used in place of their parent row 77. 
\end{tablenotes}
\end{threeparttable}
\end{table}

Neighboring municipalities are identified using first-order queen contiguity on the fixed 2020 municipal map, where two municipalities are considered neighbors if their geographic polygons share a boundary segment or a boundary point, including across prefectural borders. 
Neighboring supply is constructed using all 1,741 municipalities in the spatial source data before restricting the estimation sample to 1,725 municipalities with complete observations. 
The balanced estimation panel has 1,725 municipalities and three observations per municipality, totaling 5,175 municipality-wave observations. 
The 2012 Census excluded seven Fukushima municipalities completely and four partially \citep{fukushima2012ecensus}. Missing local or neighboring establishment counts reduce the balanced sample from 1,741 to 1,725 municipalities. 
The 2021 Census expanded its establishment register using corporate-number records \citep{micmeti2023ecensus}. 

\subsection{Local and Neighboring Food Retail}\label{subsec:local_neighboring_retail}

Let $P_{m \tau}$ be resident population in municipality $m$ and Census wave $\tau$. 
Denote $R_{m \tau}$ and $x_{m \tau} = \log(1 + R_{m \tau})$ the indexed-by-wave version of $R_m$ and $x_m$, respectively. 
Define 
\begin{equation}
\label{eq:population_variable}
Y_{m \tau} \coloneqq \log P_{m \tau}. 
\end{equation}
Since all observations are harmonized to fixed 2020 municipal boundaries, municipal area is time-invariant. 
Thus, log population and log population density differ only by a municipality-specific constant, which is absorbed by municipality fixed effects. 

Let $W = (w_{mj})$ be the fixed, row-standardized queen-contiguity matrix.
If municipality $m$ has $k_m > 0$ queen neighbors, $w_{mj} = 1 / k_m$ for each such neighbor $j$ and $w_{mj}=0$ otherwise. 
If $k_m = 0$, the entire row is zero. 
Thus $w_{mm}=0$, and the row sum is one whenever $m$ has at least one neighbor. 
Define the neighboring food retail as 
\begin{equation}
\label{eq:neighboring_retail}
e_{m \tau} \coloneqq \sum_j w_{mj} x_{j \tau}. 
\end{equation}
This variable averages transformed establishment counts in neighboring municipalities. 

The main regression specification is 
\begin{equation}
\label{eq:twfe_slx}
Y_{m \tau} = \alpha_m + \lambda_\tau + \beta x_{m \tau} + \theta e_{m \tau} + u_{m \tau}, 
\end{equation}
where $\alpha_m$ and $\lambda_\tau$ are municipality and wave fixed effects, respectively. 
This is a spatial lag of X (SLX) specification \citep{lesage2009introduction,halleckvega2015slx}. 
The coefficients $\beta$ and $\theta$ describe conditional within-municipality associations with local and neighboring FRS, respectively. 
Because retail entry also responds to population, these coefficients do not separately identify the structural responses derived in Section \ref{sec:framework}. 
In particular, $\theta$ differs from the total equilibrium response in Equation \eqref{eq:neighboring_population_response}, which allows local retail supply to adjust to changes in neighboring supply. 

\subsection{Variation across Retail Scale}\label{subsec:retail_scale}

To allow the local slope to vary, I estimate
\begin{equation}
\label{eq:quadratic_retail}
Y_{m \tau} = \alpha_m + \lambda_\tau + \beta_c(x_{m \tau} - c) + \frac{\kappa}{2}(x_{m \tau} - c)^2 + \theta_c e_{m \tau} + u_{m \tau}, 
\end{equation}
where $c$ is the pooled sample median of transformed local FRS. 
Thus, $\beta_c$ represents the fitted local slope at the median. 
The slope of the fitted relationship with respect to transformed local FRS is 
\begin{equation}
\label{eq:local_slope}
b(x) = \beta_c + \kappa(x - c). 
\end{equation}
A negative $\kappa$ means that the fitted slope decreases with establishment counts. 

To assess sensitivity to functional form, I also consider restricted cubic splines with four knots fixed at the pooled 5th, 35th, 65th, and 95th percentiles of transformed local FRS. 
The nonlinear specifications are compared under common wave effects and prefecture-by-wave effects. 
I further examine their sensitivity to excluding the final paired observation wave. 

\subsection{Food Retail and Food Service}\label{subsec:frs_fss_comparison}

Food retail and food service provide different ways to obtain food. 
Their relationships with resident population may differ because meal preparation, convenience, and the customers they serve differ. 
To compare them, let $S_{m \tau}$ denote the number of FSS establishments, $s_{m \tau}=\log(1 + S_{m \tau})$, and $e^S_{m \tau}=\sum_j w_{mj}s_{j \tau}$. 
I estimate the following joint specification on the same balanced municipality panel used in Equation \eqref{eq:twfe_slx}: 
\begin{equation}
\label{eq:frs_fss_joint}
Y_{m \tau} = \alpha_m + \lambda_\tau + b_R x_{m \tau} + b_S s_{m \tau} + \theta_R e_{m \tau} + \theta_S e^S_{m \tau} + u_{m \tau}. 
\end{equation}
The local FRS coefficient now conditions on both local FSS and the neighboring measures of the two sectors. 

For a comparison at each sector's observed scale of variation, define 
\begin{equation}
\label{eq:frs_fss_contrast}
\Delta_{RS} = b_R \sigma_x^w - b_S \sigma_s^w, 
\end{equation}
where $\sigma_x^w$ and $\sigma_s^w$ are the standard deviations of $x$ and $s$ after removing the same fixed effects on the common sample, respectively. 
Each term measures the fitted difference in log population associated with one standard deviation of the corresponding residualized supply variable. 

Statistical inference uses a Conley-type spatial heteroskedasticity and autocorrelation consistent (HAC) covariance estimator \citep{conley1999gmm}. 
The primary spatial bandwidth is 100 km, with 200 km used for sensitivity analysis. 
The covariance estimator applies a uniform distance kernel to municipal representative points. 
Scores are summed across the observed waves within each municipality before calculating spatial covariance, allowing cross-wave error dependence both within and between nearby municipalities. 

\section{Empirical Findings} \label{sec:empirical_facts}

\subsection{Local and Neighboring FRS} \label{subsec:empirical_local_neighbor}

Table \ref{tab:linear_frs} reports the linear population--FRS associations. 
Column (1) includes local FRS with municipality and wave fixed effects. 
The estimated coefficient is 0.1065, indicating a positive within-municipality association between resident population and local food retail supply. 

\begin{figure}[!t]
\centering
\includegraphics[width=\textwidth]{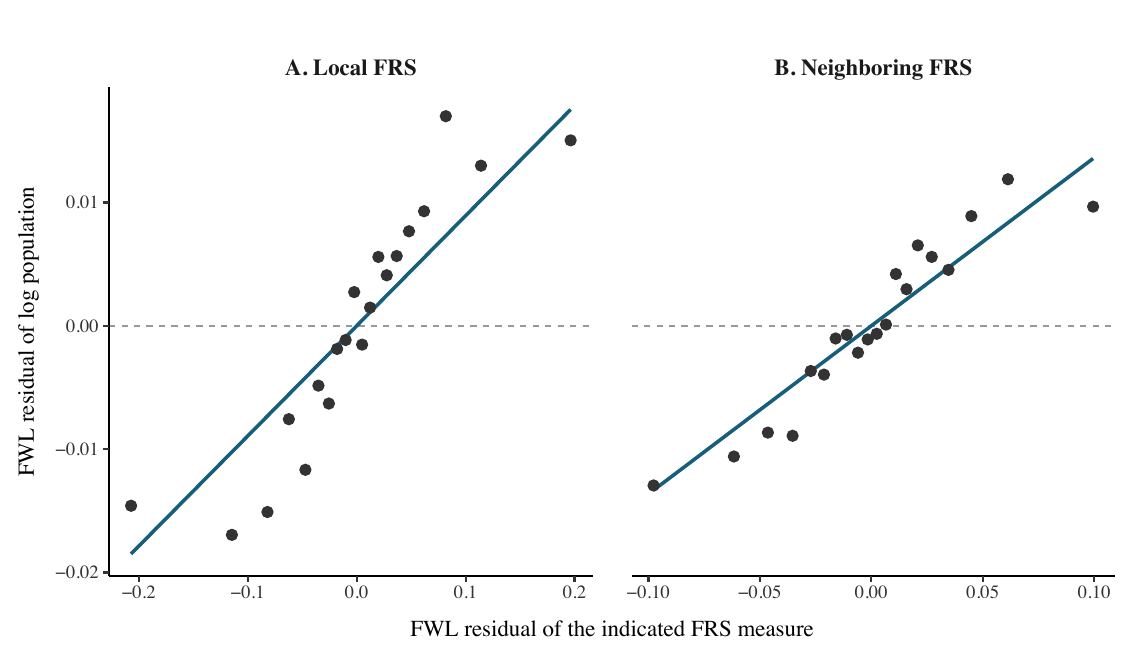}
\caption{
Conditional local and neighboring FRS associations with log population. 
Both panels use the specification in column (2) of Table \ref{tab:linear_frs}. 
Panel A displays local FRS, and Panel B displays neighboring FRS. 
The horizontal and vertical axes show Frisch–Waugh–Lovell (FWL) residuals after removing municipality and wave fixed effects and the other FRS measure. 
Points represent means within 20 quantile bins of the residualized horizontal variable. 
Solid lines correspond to the full-sample regression coefficients, with a zero intercept in residual space. 
}
\label{fig:conditional_frs}
\end{figure}

Column (2) adds neighboring FRS and serves as the main empirical specification. 
The local FRS coefficient is 0.0893, while the neighboring coefficient is 0.1364. 
Both estimates are positive, with 95\% confidence intervals excluding zero. 
Figure \ref{fig:conditional_frs} illustrates these relationships after removing fixed effects and the other supply variable using Frisch--Waugh--Lovell residualization. 

Column (3) replaces common wave effects with prefecture-by-wave fixed effects. 
The local and neighboring coefficients remain positive at 0.0904 and 0.1291, respectively. 
Their limited changes suggest that the average associations are not driven entirely by time variation common to municipalities within the same prefecture. 
Using distance-weighted five-nearest-neighbor links gives local and neighboring coefficients of 0.0839 and 0.1647. 
Allowing 2010 population density to interact with wave indicators alongside prefecture-by-wave effects reduces the local coefficient from 0.0904 to 0.0687 and the neighboring coefficient from 0.1291 to 0.0765. 
The larger decline in the neighboring coefficient indicates that this association is more sensitive to population trends related to initial density. 
Pairing retail supply in 2012 and 2016 with subsequent population in 2015 and 2020 gives 0.0872 and 0.1349 under prefecture-by-wave effects. 
When the original paired sample is restricted to the first two waves, the corresponding coefficients are 0.0616 and 0.1198. 
The local estimate has a 95\% confidence interval of [$-0.0002$, 0.1233]. 
At the median of 69 establishments, an increase to 76 establishments corresponds to approximately 0.85\% higher fitted population using the column (2) coefficient. 

Excluding 27 municipalities in Iwate and Miyagi with documented tsunami inundation in 2011 \citep{gsi2011tsunami} yields local and neighboring FRS coefficients of 0.1114 and 0.1530 under wave fixed effects, and 0.1134 and 0.1556 under prefecture-by-wave fixed effects. 
The same exclusion is also applied to the joint FRS–FSS specification discussed below. 

\begin{table}[!t]
\centering
\begin{threeparttable}
\caption{Linear population associations with local and neighboring FRS}
\label{tab:linear_frs}
\normalsize
\setlength{\tabcolsep}{4pt}
\renewcommand{\arraystretch}{1.15}
\begin{tabularx}{\textwidth}{Yccc}
\toprule
 & (1) & (2) & (3) \\
\midrule
Local FRS, $x=\log(1+R)$ & 0.1065 & 0.0893 & 0.0904 \\
 & [0.0540, 0.1590] & [0.0422, 0.1364] & [0.0428, 0.1381] \\
\addlinespace
Neighboring FRS, $e=Wx$ & --- & 0.1364 & 0.1291 \\
 &  & [0.0597, 0.2131] & [0.0459, 0.2123] \\
\addlinespace
\midrule
Municipality fixed effects & Yes & Yes & Yes \\
Wave fixed effects & Yes & Yes & --- \\
Prefecture $\times$ wave fixed effects & --- & --- & Yes \\
Municipalities & 1,725 & 1,725 & 1,725 \\
Observations & 5,175 & 5,175 & 5,175 \\
\bottomrule
\end{tabularx}
\begin{tablenotes}[flushleft]
\small
\item \textit{Notes:}
The dependent variable is log resident population. 
The balanced sample contains 1,725 municipalities (2020 boundaries) and 5,175 paired observations. 
Column (1) includes local FRS only. 
Column (2) adds neighboring FRS. 
Column (3) replaces wave effects with prefecture-by-wave effects. 
FRS denotes broad food and beverage retail establishments. 
Brackets report 95\% normal intervals using a uniform-kernel spatial HAC covariance with a 100 km bandwidth and municipality scores summed across all waves. 
\end{tablenotes}
\end{threeparttable}
\end{table}

\subsection{FRS and FSS}\label{subsec:empirical_frs_fss}

Table \ref{tab:frs_fss} reports the joint FRS--FSS specification using the same municipalities and fixed effects as the main FRS regression. 
The estimated local FRS coefficient is 0.0928, with a 95\% confidence interval of [0.0507, 0.1350]. 
The corresponding local FSS coefficient is $-0.0052$, with a confidence interval of [$-0.0467$, 0.0363]. 

To compare the two sectors on comparable scales, I use the standardized difference $\Delta_{RS}$ defined in Equation \eqref{eq:frs_fss_contrast}. 
The estimated difference is 0.0088, with a 95\% confidence interval of [0.0039, 0.0138]. 
The interval excludes zero, indicating a stronger standardized local FRS association than FSS.

\begin{figure}[!t]
\centering
\includegraphics[width=\textwidth]{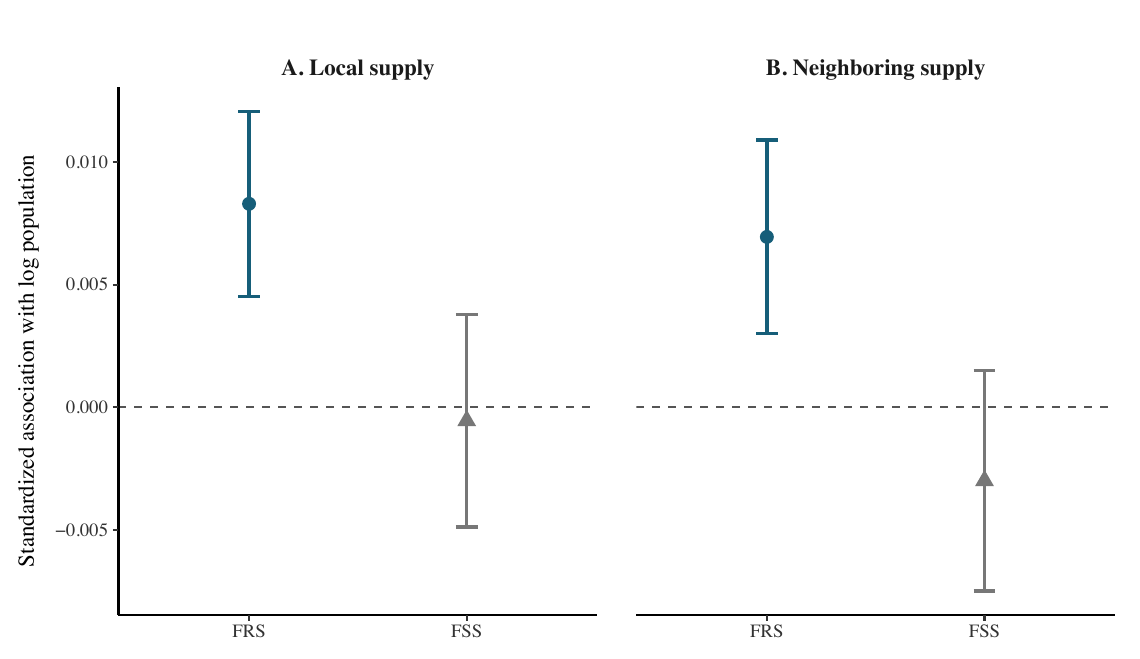}
\caption{
Standardized local and neighboring FRS–FSS associations with log population. 
Estimates come from the joint FRS–FSS specification in Table \ref{tab:frs_fss}, using 1,725 municipalities and 5,175 observations. 
Panel A reports local supply, and Panel B reports neighboring supply. Each point represents the estimated coefficient multiplied by the standard deviation of the corresponding supply variable after removing municipality and wave fixed effects. 
Circles denote FRS, and triangles denote FSS. 
Vertical bars report 95\% confidence intervals based on the 100 km spatial HAC estimator. The dashed horizontal line indicates zero. 
}
\label{fig:standardized_frs_fss}
\end{figure}

Figure \ref{fig:standardized_frs_fss} illustrates the four associations on comparable standardized scales. 
Neighboring FRS is positively associated with population, while the neighboring FSS estimate is imprecise. 
The standardized local difference remains positive with prefecture-by-wave fixed effects and after removing the final paired wave. 
Table \ref{tab:frs_fss_robustness} reports four joint specifications. 
The estimated difference ranges from 0.0055 to 0.0088, with all 95\% intervals excluding zero. 

After excluding the same 27 municipalities, the standardized local FRS--FSS difference is 0.0074 with wave fixed effects and 0.0065 with prefecture-by-wave fixed effects. 
Both 95\% intervals exclude zero. 
The local FSS coefficient under prefecture-by-wave fixed effects is 0.0280, with a 95\% interval of [0.0054, 0.0506]. 
Table \ref{tab:tsunami_sensitivity} reports the results of both specifications after excluding tsunami-inundated municipalities. 

\begin{table}[!t]
\centering
\begin{threeparttable}
\caption{Population associations with local and neighboring FRS and FSS}
\label{tab:frs_fss}
\normalsize
\setlength{\tabcolsep}{4pt}
\renewcommand{\arraystretch}{1.15}
\begin{tabularx}{\textwidth}{Ycc}
\toprule
Parameter or contrast & Estimate & 95\% interval \\
\midrule
Local FRS, $x$: $b_R$ & 0.0928 & [0.0507, 0.1350] \\
Local FSS, $s$: $b_S$ & -0.0052 & [-0.0467, 0.0363] \\
Neighboring FRS, $e$: $\theta_R$ & 0.1543 & [0.0667, 0.2420] \\
Neighboring FSS, $e^S$: $\theta_S$ & -0.0600 & [-0.1503, 0.0303] \\
\midrule
Within SD of local FRS, $\sigma_x^w$ & 0.089346 & --- \\
Within SD of local FSS, $\sigma_s^w$ & 0.104553 & --- \\
Standardized local difference, $\Delta_{RS}$ & 0.0088 & [0.0039, 0.0138] \\
\midrule
Municipality and wave fixed effects & \multicolumn{2}{c}{Yes} \\
Municipalities / observations & \multicolumn{2}{c}{1,725 / 5,175} \\
\bottomrule
\end{tabularx}
\begin{tablenotes}[flushleft]
\small
\item \textit{Notes:}
The dependent variable is log resident population. 
The balanced sample contains 1,725 municipalities (2020 boundaries) and 5,175 paired observations. 
Population in 2010, 2015, and 2020 precedes supply in 2012, 2016, and 2021, respectively. 
Brackets report 95\% normal intervals using a uniform-kernel spatial HAC covariance with a 100 km bandwidth and municipality scores summed across all waves. 
$\Delta_{RS} = b_R\sigma_x^w - b_S\sigma_s^w$ compares one within-sample SD of each local sector. 
\end{tablenotes}
\end{threeparttable}
\end{table}

\begin{table}[t]
\centering
\caption{Robustness of standardized local FRS--FSS differences}
\label{tab:frs_fss_robustness}

\begingroup
\normalsize
\setlength{\tabcolsep}{3pt}
\begin{tabular*}{\linewidth}{@{\extracolsep{\fill}}lcccc@{}}
\toprule
Specification & $b_R$ & $b_S$ & $\Delta_{RS}$ & 95\% interval \\
\midrule
Wave FE, full & 0.0928 & $-0.0052$ & 0.0088 & $[0.0039,0.0138]$ \\
Pref.$\times$wave FE, full & 0.0917 & 0.0024 & 0.0076 & $[0.0031,0.0122]$ \\
Wave FE, early & 0.0677 & $-0.0177$ & 0.0062 & $[0.0031,0.0094]$ \\
Pref.$\times$wave FE, early & 0.0656 & $-0.0129$ & 0.0055 & $[0.0026,0.0084]$ \\
\bottomrule
\end{tabular*}

\par\vspace{2pt}
\noindent
\parbox{\linewidth}{
\textit{Notes:} All specifications contain 1,725 municipality fixed effects. 
``Early'' excludes the final paired wave (3,450 versus 5,175 observations). 
Intervals use 100 km spatial HAC. 
$\Delta_{RS}$ uses within-specification standard deviations.
}
\endgroup
\end{table}

\begin{table}[t]
\centering
\begin{threeparttable}
\caption{Population associations excluding tsunami-inundated municipalities}
\label{tab:tsunami_sensitivity}
\normalsize
\setlength{\tabcolsep}{4pt}
\renewcommand{\arraystretch}{1.15}
\begin{tabularx}{\textwidth}{Ycc}
\toprule
 & Wave FE & Prefecture $\times$ wave FE \\
\midrule
\multicolumn{3}{@{}l}{Panel A. Local and neighboring FRS} \\
Local FRS, $x$ & 0.1114 & 0.1134 \\
 & [0.0743, 0.1486] & [0.0768, 0.1501] \\
\addlinespace
Neighboring FRS, $e$ & 0.1530 & 0.1556 \\
 & [0.0735, 0.2324] & [0.0753, 0.2359] \\
\midrule
\multicolumn{3}{@{}l}{Panel B. Joint FRS--FSS specification} \\
Local FRS, $x$ & 0.1091 & 0.1099 \\
 & [0.0727, 0.1455] & [0.0742, 0.1455] \\
\addlinespace
Local FSS, $s$ & 0.0209 & 0.0280 \\
 & [$-0.0047$, 0.0466] & [0.0054, 0.0506] \\
\addlinespace
Neighboring FRS, $e$ & 0.1563 & 0.1552 \\
 & [0.0659, 0.2467] & [0.0691, 0.2413] \\
\addlinespace
Neighboring FSS, $e^S$ & $-0.0224$ & $-0.0132$ \\
 & [$-0.1104$, 0.0655] & [$-0.0657$, 0.0392] \\
\midrule
Standardized local difference, $\Delta_{RS}$ & 0.0074 & 0.0065 \\
 & [0.0031, 0.0117] & [0.0024, 0.0105] \\
\midrule
Municipalities & 1,698 & 1,698 \\
Observations & 5,094 & 5,094 \\
\bottomrule
\end{tabularx}
\begin{tablenotes}[flushleft]
\small
\item \textit{Notes:}
The same 27 Iwate and Miyagi municipalities are excluded in both panels.
All specifications include municipality fixed effects and retain the original neighboring-supply measures.
Brackets give 95\% intervals using 100 km spatial HAC.
$\Delta_{RS}$ uses the within-sample standard deviations under each fixed-effects specification.
\end{tablenotes}
\end{threeparttable}
\end{table}

\subsection{Variation in the Local FRS Relationship}\label{subsec:empirical_nonlinearity}

\begin{figure}[!t]
\centering
\includegraphics[width=\textwidth]{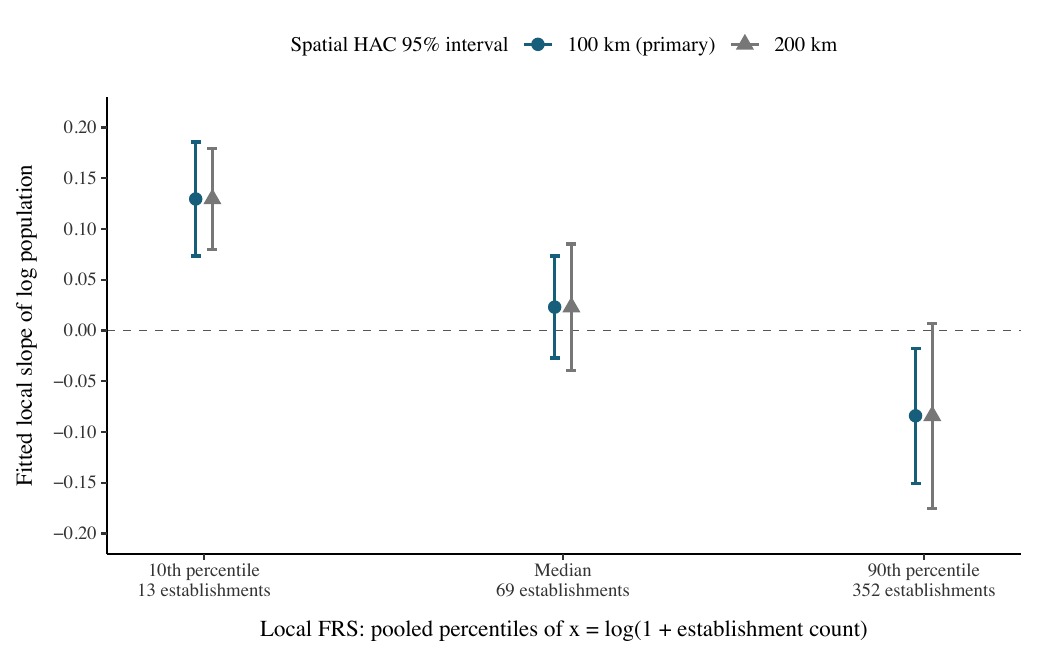}
\caption{
Fitted local FRS slopes across retail scale. 
Estimates come from the benchmark quadratic TWFE–SLX specification in Table \ref{tab:quadratic_frs}, using 1,725 municipalities and 5,175 observations. 
Points report the fitted local slopes at the pooled 10th percentile, median, and 90th percentile of transformed local FRS. 
Circles and triangles display the same point estimates with 100 km and 200 km spatial HAC confidence intervals, respectively. 
Vertical bars report pointwise 95\% confidence intervals accounting for the joint covariance of the linear and quadratic coefficients. The dashed horizontal line indicates zero. 
}
\label{fig:local_frs_slope}
\end{figure}

Table \ref{tab:quadratic_frs} reports the benchmark quadratic specification, and Figure \ref{fig:local_frs_slope} illustrates its fitted local slopes and confidence intervals. 
The estimated quadratic coefficient is $-0.0662$, indicating a declining fitted slope under municipality and common wave fixed effects. 
The slope is 0.1296 at the 10th percentile of FRS, 0.0230 at the median, and $-0.0841$ at the 90th percentile. 
The median slope is imprecisely estimated, while the negative 90th-percentile slope excludes zero at the primary 100 km bandwidth but not at 200 km. 

The nonlinear results are more sensitive to regional time effects than the linear associations. 
When prefecture-by-wave fixed effects replace common wave effects, the quadratic coefficient changes from $-0.0662$ to $-0.0338$. 
The fitted slope at the 90th percentile increases from $-0.0841$ to approximately zero. 
The difference between the fitted slopes at the 90th and 10th percentiles remains negative, but its magnitude declines from $-0.2137$ to $-0.1091$. 
Thus, the full-sample quadratic evidence supports a weaker conditional association at higher retail scale. 

The four-knot restricted cubic spline also indicates heterogeneity across retail scale. 
Its fitted slope at the 90th percentile changes from $-0.1870$ under common wave effects to $-0.0889$ under prefecture-by-wave effects. 
The corresponding confidence intervals remain negative in the full three-wave sample. 
However, the estimated slope around the median differs substantially from the quadratic specification. 
Thus, we cannot conclude that the local slope approaches zero at the median or declines monotonically throughout the observed distribution. 

The nonlinear relationship also depends on the observation periods. 
Excluding the final paired wave reduces the quadratic coefficient from $-0.0662$ to $-0.0236$ under common wave effects. 
When prefecture-by-wave effects are also included, the coefficient becomes $-0.0023$, and the fitted 90th-percentile slope is 0.0554. 
The corresponding four-knot spline slope at the 90th percentile is $-0.0216$ and is imprecisely estimated. 
Thus, the evidence supports only specification-dependent heterogeneity across retail scale. 
Table \ref{tab:nonlinear_robustness} summarizes these robustness check results. 

\begin{table}[!t]
\centering
\begin{threeparttable}
\caption{Quadratic population–FRS associations and fitted local slopes}
\label{tab:quadratic_frs}
\normalsize
\setlength{\tabcolsep}{4pt}
\renewcommand{\arraystretch}{1.15}
\begin{tabularx}{\textwidth}{Ycc}
\toprule
Parameter or fitted slope & Estimate & 95\% interval \\
\midrule
\multicolumn{3}{l}{\textit{Panel A. Quadratic TWFE--SLX}} \\
Centered local FRS, $x - c$: $\beta_c$ & 0.0230 & [-0.0272, 0.0733] \\
Half-square, $(x - c)^2 / 2$: $\kappa$ & -0.0662 & [-0.0883, -0.0442] \\
Neighboring FRS, $e$: $\theta_c$ & 0.1595 & [0.0872, 0.2318] \\
\addlinespace
\multicolumn{3}{l}{\textit{Panel B. Local slope with respect to $x = \log(1 + R)$}} \\
10th percentile (13 establishments) & 0.1296 & [0.0734, 0.1858] \\
Median (69 establishments) & 0.0230 & [-0.0272, 0.0733] \\
90th percentile (352 establishments) & -0.0841 & [-0.1507, -0.0175] \\
\midrule
Municipality and wave fixed effects & \multicolumn{2}{c}{Yes} \\
Municipalities / observations & \multicolumn{2}{c}{1,725 / 5,175} \\
\bottomrule
\end{tabularx}
\begin{tablenotes}[flushleft]
\small
\item \textit{Notes:}
The dependent variable is log resident population. 
The balanced sample contains 1,725 municipalities (2020 boundaries) and 5,175 paired observations. 
Population in 2010, 2015, and 2020 precedes supply in 2012, 2016, and 2021, respectively. 
Brackets report 95\% normal intervals using a uniform-kernel spatial HAC covariance with a 100 km bandwidth and municipality scores summed across all waves. 
The center is the pooled sample median, $c=4.248495$. 
Slopes are $b(x) = \beta_c + \kappa(x - c)$. 
\end{tablenotes}
\end{threeparttable}
\end{table}

\begin{table}[t]
\centering
\caption{Robustness of fitted local FRS slopes}
\label{tab:nonlinear_robustness}

\begingroup
\normalsize
\setlength{\tabcolsep}{3pt}

\begin{tabular}{@{}lcccc@{}}
\toprule
Specification
& $\hat b(P_{10})$
& $\hat b(P_{50})$
& $\hat b(P_{90})$
& $\Delta_{90,10}$ \\
\midrule

\multicolumn{5}{@{}l}{\textit{Panel A. Quadratic}} \\

Wave FE, full
& 0.1296
& 0.0230
& $-0.0841$
& $-0.2137$ \\

& \shortstack[c]{$[0.0734,$\\$\quad 0.1858]$}
& \shortstack[c]{$[-0.0272,$\\$\quad 0.0733]$}
& \shortstack[c]{$[-0.1507,$\\$\quad -0.0175]$}
& \shortstack[c]{$[-0.2849,$\\$\quad -0.1425]$} \\

Pref.$\times$wave FE, full
& 0.1093
& 0.0548
& 0.0001
& $-0.1091$ \\

& \shortstack[c]{$[0.0590,$\\$\quad 0.1595]$}
& \shortstack[c]{$[0.0031,$\\$\quad 0.1066]$}
& \shortstack[c]{$[-0.0619,$\\$\quad 0.0622]$}
& \shortstack[c]{$[-0.1542,$\\$\quad -0.0641]$} \\

Wave FE, early
& 0.0759
& 0.0379
& $-0.0003$
& $-0.0762$ \\

& \shortstack[c]{$[0.0107,$\\$\quad 0.1411]$}
& \shortstack[c]{$[-0.0302,$\\$\quad 0.1060]$}
& \shortstack[c]{$[-0.0852,$\\$\quad 0.0846]$}
& \shortstack[c]{$[-0.1422,$\\$\quad -0.0102]$} \\

Pref.$\times$wave FE, early
& 0.0628
& 0.0591
& 0.0554
& $-0.0075$ \\

& \shortstack[c]{$[0.0004,$\\$\quad 0.1253]$}
& \shortstack[c]{$[-0.0080,$\\$\quad 0.1262]$}
& \shortstack[c]{$[-0.0326,$\\$\quad 0.1433]$}
& \shortstack[c]{$[-0.0800,$\\$\quad 0.0650]$} \\

\midrule

\multicolumn{5}{@{}l}{\textit{Panel B. Restricted cubic spline}} \\

Wave FE, full
& 0.1049
& 0.1043
& $-0.1870$
& $-0.2919$ \\

& \shortstack[c]{$[0.0504,$\\$\quad 0.1594]$}
& \shortstack[c]{$[0.0441,$\\$\quad 0.1645]$}
& \shortstack[c]{$[-0.2611,$\\$\quad -0.1130]$}
& \shortstack[c]{$[-0.3746,$\\$\quad -0.2093]$} \\

Pref.$\times$wave FE, full
& 0.0849
& 0.1246
& $-0.0889$
& $-0.1739$ \\

& \shortstack[c]{$[0.0330,$\\$\quad 0.1369]$}
& \shortstack[c]{$[0.0705,$\\$\quad 0.1788]$}
& \shortstack[c]{$[-0.1638,$\\$\quad -0.0140]$}
& \shortstack[c]{$[-0.2335,$\\$\quad -0.1142]$} \\

Wave FE, early
& 0.0565
& 0.0945
& $-0.1111$
& $-0.1676$ \\

& \shortstack[c]{$[-0.0197,$\\$\quad 0.1328]$}
& \shortstack[c]{$[0.0183,$\\$\quad 0.1707]$}
& \shortstack[c]{$[-0.1964,$\\$\quad -0.0258]$}
& \shortstack[c]{$[-0.2620,$\\$\quad -0.0732]$} \\

Pref.$\times$wave FE, early
& 0.0434
& 0.1051
& $-0.0216$
& $-0.0650$ \\

& \shortstack[c]{$[-0.0299,$\\$\quad 0.1167]$}
& \shortstack[c]{$[0.0329,$\\$\quad 0.1772]$}
& \shortstack[c]{$[-0.1152,$\\$\quad 0.0720]$}
& \shortstack[c]{$[-0.1623,$\\$\quad 0.0322]$} \\

\bottomrule
\end{tabular}

\par\vspace{2pt}
\noindent
\parbox{\linewidth}{
\textit{Notes:}
Brackets give 95\% intervals (100 km spatial HAC).
$\Delta_{90,10} = \hat b(P_{90}) - \hat b(P_{10})$.
``Early'' excludes the final paired wave.
}

\endgroup
\end{table}

\section{Conclusion}\label{sec:conclusion}

Food retail supply and resident population are jointly organized through local demand and consumption amenities. 
The theoretical framework shows how retail access, housing adjustment, and retail entry can generate positive local feedback, while neighboring establishments connect retail markets across municipal boundaries. 

The empirical analysis provides two main findings. 
First, local and neighboring food retail supply are positively associated with resident population. 
These relationships remain positive under alternative spatial weights and baseline-density trend controls. 
Second, the standardized local association is stronger for food retail than for food service in a common-sample specification. 
This difference persists with prefecture-by-wave fixed effects and after excluding the final paired wave. 

The nonlinear analysis indicates that the local FRS association varies across retail scale. 
The estimated curvature is sensitive to prefecture-specific time variation, functional form, and the observation period. 

The results indicate that local food retail and resident population are closely related within and across municipal boundaries. 
They also show why establishment counts should be interpreted in their local market context, rather than treated as complete measures of food access. 
These findings highlight the importance of local and neighboring retail markets in understanding the spatial organization of essential consumption services.

\section*{Declaration on the Use of AI Tools}

OpenAI's ChatGPT and Codex were used as supporting tools during the preparation and verification of this manuscript in 2026.
Their assistance comprised four tasks.
First, they reviewed grammar, typographical errors, readability, and the accuracy of written expression.
Second, under the research design and instructions established by the author, they assisted in retrieving publicly available official data and checking data provenance, definitions, completeness, and consistency.
Third, they helped compare the empirical results, tables, figures, and corresponding manuscript descriptions with the analysis outputs and code to identify possible discrepancies.
Fourth, under the author's specific instructions, they assisted in adjusting figure layouts, font sizes, and visual presentation.
The author reviewed and validated all analytical code, including code developed or revised with AI assistance, and verified the data processing, empirical results, figures, and manuscript text.

All research questions, theoretical and empirical decisions, interpretations, and final revisions remained under the author's control.
The author takes full responsibility for the accuracy, integrity, and reproducibility of the data processing, code, empirical results, figures, and submitted manuscript, including any errors that may arise.

\clearpage
\newpage

\bibliographystyle{plainnat}
\bibliography{references}

\end{document}